\documentclass[twocolumn,notitlepage,superscriptaddress,amssymb, nobibnotes,aps]{revtex4-1}

\expandafter\let\csname equation*\endcsname\relax
\expandafter\let\csname endequation*\endcsname\relax
\expandafter\let\csname eqnarray*\endcsname\relax
\expandafter\let\csname endeqnarray*\endcsname\relax
\usepackage{amsmath}
\usepackage{amssymb}
\usepackage{graphicx}

\newcommand{\RN}[1]{\textup{\uppercase\expandafter{\romannumeral#1}}}

\renewcommand{\rm}{\mathrm}

\usepackage[hidelinks]{hyperref}

\begin{document}
\title{Testing small scale gravitational wave detectors with dynamical mass distributions}

\author{Dennis R\"atzel}
\email{dennis.raetzel@univie.ac.at}
\affiliation{Faculty of Physics, University of Vienna, Boltzmanngasse 5, 1090 Vienna, Austria}

\author{Ivette Fuentes}
\affiliation{School of Mathematical Sciences, University of Nottingham, University Park, Nottingham NG7 2RD, UK}
\affiliation{Faculty of Physics, University of Vienna, Boltzmanngasse 5, 1090 Vienna, Austria}

\begin{abstract}
The recent discovery of gravitational waves by the LIGO-Virgo collaboration created renewed interest in the investigation of alternative gravitational detector designs, such as small scale resonant detectors. In this article, it is shown how proposed small scale detectors can be tested by generating dynamical gravitational fields with appropriate distributions of moving masses. A series of interesting experiments will be possible with this setup. In particular, small scale detectors can be tested very early in the development phase and tests can be used to progress quickly in their development. This could contribute to the emerging field of gravitational wave astronomy.\\


\end{abstract}

\maketitle

\section{Introduction}
The first observation of gravitational waves by the LIGO-Virgo collaboration in November 2015 \cite{Abbott:2016obs} is seen as the dawn of the age of gravitational wave astronomy. In the next few years, new interferometric detectors will be built and developed, enabling us to learn more about the sources of gravitational waves and the cosmos. The success of the LIGO-Virgo collaboration also sparked renewed interest into the development of alternative detector designs on much smaller scales than the kilometers required for an interferometric detector. Among such proposals are some that consider detectors on scales between meters and micrometers, we shall refer to these as small scale detectors. Proposed systems range from electromagnetic cavity resonators \cite{Caves:1979mic,Reece:1982sc,Gemme:2001ba,Ballatini:2003ade} and resonant mass detectors \cite{Ju:2000det,Maggiore:2008grav} over Bose-Einstein condensates \cite{Sabin:2014bua,Sabin:2015mha} 
to a microwave cavity resonator coupled to a superfluid helium container \cite{Singh:2016xwa,DeLorenzo:2016ult}. Some proposals for small scale detectors are based on quantum technologies, for example the one presented in \cite{Sabin:2014bua,Sabin:2015mha}. With the advancement of quantum technologies that will be promoted by funding initiatives like that by the European Commission in the next years, we expect many more proposals for small scale gravitational wave detectors to appear. 

The main advantage of small scale detectors is that a single detector would be cheap in comparison to an interferometric detector and a lot of them could be built to achieve high directional resolution. One could even imagine a network of hundreds of small scale gravitational wave detectors throughout the world with collective data analysis. The disadvantages of small scale detectors are: they are usually narrow band, they often operate in a high frequency band and they need long integration times to achieve the necessary sensitivity for a detection. This makes most of them only applicable for persistent sources of gravitational waves. The disadvantages could be possibly overcome by design adjustments and long term development. In particular, it would be beneficial if prototypes could be tested and evaluated with artificially created gravitational signals of larger amplitudes than those expected from gravitational wave sources. Unfortunately, it is nearly impossible to artificially create gravitational waves of significant amplitude. However, the effect of local oscillating gravitational fields on a small scale detector can resemble the effect of gravitational waves sufficiently to serve as test signals which can be created with comparatively large amplitudes. In this letter, we propose tests of small scale gravitational wave detectors that could be performed by employing moving masses, creating local gravitational fields that resemble gravitational waves on the length scale of the detectors. 

The basic framework that we will employ in this letter is linearized gravity, where the spacetime metric $g_{\mu\nu}$ is considered to differ just slightly from the flat Minkowski metric $\eta_{\mu\nu}$. We define the perturbation of the metric as $h_{\mu\nu}=g_{\mu\nu}-\eta_{\mu\nu}$ and we assume that $|h_{\mu\nu}|\ll 1$ for all $\mu,\nu$ in an appropriately chosen set of coordinates. 

\section{Gravitational wave metric} Since sources of gravitational waves are usually very distant, we can restrict our considerations to plane gravitational waves.  The perturbation of the spacetime metric $h^\rm{w}_{\mu\nu}$ (the superscript w standing for wave) corresponding to a plane gravitational wave that propagates in the positive or negative $z$-direction has only four non-zero components $h^\rm{w}_{11}=-h^\rm{w}_{22}=h_+$ and $h^\rm{w}_{12}=h^\rm{w}_{21}=h_\times$ in the transversal-traceless (TT) gauge which is a specific coordinate system in which freely falling massive test particle at rest stay at rest. Here, $h_+$ and $h_\times$ are the strains corresponding to the two polarization directions $+$ and $\times$, respectively. If we assume propagation in the positive $z$-direction and consider a monochromatic wave of angular frequency $\omega$, the corresponding strains can be written as 
\begin{eqnarray}\label{eq:strains}
h_{+,\omega}(t,z)&=&h_{0,+,\omega}\cos(\omega (t - z) + \varphi_+)\\
\nonumber h_{\times,\omega}(t,z)&=&h_{0,\times,\omega}\cos(\omega (t - z) + \varphi_\times)\,.
\end{eqnarray}

The TT gauge corresponds to an appropriate coordinate system for the analysis of interferometric detectors such as  LIGO and Virgo, since their mirrors are, in approximation, freely falling along the beam line of the interferometer. For the analysis of local detectors, a different coordinate system is much better suited - the proper detector frame \cite{Maggiore:2008grav}, which was introduced in \cite{Ni1978proper}. The proper detector frame corresponds to the coordinate system constructed by an observer using rigid rods to define spatial coordinates from the center of the observer's laboratory and a clock at the center of the observer's laboratory to define time. This leads to an expression of the spacetime metric in the constructed coordinates which is valid up to quadratic terms in the spatial coordinates. For a freely falling observer, the proper detector frame is commonly called Fermi normal coordinates \cite{manasse1963fermi}.

For a freely falling laboratory in the spacetime defined by $\eta_{\mu\nu}+h^\rm{w}_{\mu\nu}$, the metric perturbation in the laboratory's proper detector frame has the form \footnote{Explicit expression for a wave propagating in the negative $z$-direction can be found in \cite{Rakhmanov:2014fer}.}
\begin{equation}\label{eq:properdetM}
	h^\rm{w,P}_{\mu\nu}=\ddot h_+(t,0)M_{+,\mu\nu}/c^2 + \ddot h_\times(t,0) M_{\times,\mu\nu}/c^2\,,
\end{equation}
where the superscript P stands for proper detector frame, the dots denote the second time derivative of the strain functions and the components of the matrices $M_{+,\mu\nu}$ and $M_{\times,\mu\nu}$ are second order polynomials in $x$ and $y$ with components of order $10^{-1}$. From the geodesic equations that govern the motion of massive particles in the field of the gravitational wave, we obtain the following acceleration for any part of the small scale detector that does move with non-relativistic velocity $v^j$: 
\begin{eqnarray}\label{eq:tidalfull}
	\frac{d^2 \gamma^i}{dt^2} &\approx & \frac{c^2}{2}\left(\partial_i h^\rm{w,P}_{00} - 2\partial_t h^\rm{w,P}_{0i}/c \right.\\
	\nonumber && \left. + 2v^j\left(\partial_i h^\rm{w,P}_{0j}/c - \partial_j h^\rm{w,P}_{0i}/c - \partial_t h^\rm{w,P}_{ij}/c^2\right)\right)\,,
\end{eqnarray}
where $i,j\in\{x,y,z\}$. Eq. (\ref{eq:tidalfull}) gives the tidal forces induced by the gravitational wave that deform the small scale detector. This deformation is the fundamental mechanism on which measurement is based for all examples of small scale detectors mentioned in the introduction.

We find that the dynamical effects of all components of the metric perturbation besides $h^\rm{w,P}_{00}$ are suppressed; either by at least a factor $v/c$, where $v$ is the largest speed in the detector system (see also Sec. 17.4 of \cite{Misner1973}) or by a factor $l\omega/c$, where $l$ is the largest extension of the detector system. For all the examples of detectors given in the introduction, the detector's parts used for sensing the gravitational field are moving very slowly in comparison to the speed of light. So we can assume that $v/c\ll 1$. Furthermore, for frequencies between kHz and MHz and extensions of the detector systems at or below the meter scale, we have $l\omega/c\ll 1$. Of the various local detectors that we mentioned in the introduction, only the electromagnetic resonators \cite{Caves:1979mic,Reece:1982sc,Gemme:2001ba,Ballatini:2003ade,Singh:2016xwa,DeLorenzo:2016ult} contain parts that move at high speeds - namely the electromagnetic radiation in the resonator. However, the light is only used to "read out" deformations of the resonator, a slowly moving system of massive matter; what is used for measurement is not a direct effect of the gravitational field on the light. Therefore, this effect can be neglected \footnote{If the light is used for the sensing like in the detector proposal described in \cite{Cruise:2000ane,Cruise:2005aco,Cruise:2006apr}, the response of the detector on $g^\rm{N}_{\mu\nu}$ has to be evaluated explicitly.}.

There is also an effect on length scales by the purely spatial components of the metric perturbation; the spatial components give the length scale associated with rigid rods associated with the proper length $l_p = \int d\varsigma \sqrt{g_{\mu\nu}s^{\prime\mu} s^{\prime\nu}}$, where $s^{\prime}$ is the tangent to a space like geodesic along which the length is measured. Therefore, the change of proper length due to the metric perturbation $l_p$ is proportional to $h^\rm{w,P}_{\mu\nu}$. The deformation of the detector systems due to tidal forces, the basic mechanism of the detection process, can be derived from the acceleration $d^2 \gamma^i/dt^2$ in Eq. (\ref{eq:tidalfull}) via Hook's law. The corresponding proportionality factor is the specific modulus $Y/\rho=c_s^2$, where $Y$ is Young's modulus, $\rho$ is the mass density and $c_s$ is the speed of sound for the material the detector system consists of (see for example \cite{Ratzel:2017etl} for details). We find that the proper length change is smaller than the tidal length change by a factor $c_s^2/c^2$. The stiffest material per density is carbyne, with a specific modulus $Y/\rho$ of the order of $10^9\,\rm{m^2s^{-2}}$. The specific modulus of the used solid state matter is usually much smaller than the extreme value for carbyne. For example, the specific modulus of aluminum is $2.6\times 10^7\,\rm{m^2s^{-2}}$. This corresponds to $c_s^2/c^2 \sim 10^{-10}$. Therefore, the proper length changes can be neglected for all the detector proposals mentioned in the introduction as all realistic matter systems are far from rigid.

In summary, we are justified to restrict our considerations to the component 
\begin{equation}\label{eq:properdet}
	h^\rm{w,P}_{00}=\partial_t^2 h_+(t,0)(x^2-y^2)/2c^2 + \partial_t^2 h_\times(t,0) xy/c^2\,.
\end{equation} 
Due to our arguments above, these deformations are the only significant gravitational effects on the small scale gravitational wave detectors \footnote{This statement is also derived in \cite{Maggiore:2008grav} using the equation of geodesic deviation and the curvature tensor).}. Then, a metric perturbation that generates, to a good approximation, the same physical effects as $h^\rm{w,P}_{\mu\nu}$ can be generated by an appropriate distribution of masses, as we will show in the following. 

\section{Newtonian limit} Restricting our considerations to $h^\rm{w,P}_{00}$ and setting all other components of the metric perturbation formally to zero, we obtain a perfect Newtonian frame for the case of a gravitational wave; the spatial part of the metric is flat, diagonal and normalized and there are no space-time mixed terms in the metric (see Sec. 17.4 of \cite{Misner1973}). A similar situation arises for the mass distribution if we assume that the source masses that are used to mimic the gravitational wave metric move with non-relativistic speed. Since we already assumed that the parts of the detector used for sensing the gravitational field are moving non-relativistically, we obtain the Newtonian limit which is well defined in linearized gravity (see Sec. 6.3 of \cite{Trautman:1965lec} and Sec. 17.4 of \cite{Misner1973}). Now, we write the spacetime metric as $g^\rm{N}_{\mu\nu}= \eta_{\mu\nu} + h^\rm{N}_{\mu\nu}$. In the Newtonian limit, the only component of the metric perturbation that we have to take into account is the component $h^\rm{N}_{00}=-2\Phi/c^2$, where $\Phi$ is the Newtonian potential \cite{Trautman:1965lec,Misner1973}.  Depending on the design of the small scale gravitational wave detector, the effect of the other components of the metric perturbation on the detector may be investigated explicitly. In a specific set of coordinates (the Lorenz gauge), we can identify $h^\rm{N}_{00}=-2\Phi/c^2$ as before and 
have only three other non-zero components of the metric perturbation; the three diagonal spatial components. Furthermore, they are all equivalent to the time-time component, i.e. $h^\rm{N}_{11} = h^\rm{N}_{22} = h^\rm{N}_{33} = h^\rm{N}_{00}$ \footnote{This metric can be derived directly from the linearized Einstein equations and the energy momentum tensor for a non-relativistic point particle in the Lorenz gauge \cite{Raetzel:2016pvx}.}. However, for the detector proposals mentioned in the introduction, the same arguments that we used above to neglect all components of $h^\rm{w,P}_{\mu\nu}$ besides $h^\rm{w,P}_{00}$ can be applied to $h^\rm{N}_{ii}$. Therefore, we can restrict our considerations to $h^\rm{N}_{00}=-2\Phi/c^2$ in the following.

\section{Gravitational wave substitutes} 
\begin{figure}[h]
\includegraphics[width=6cm,angle=0]{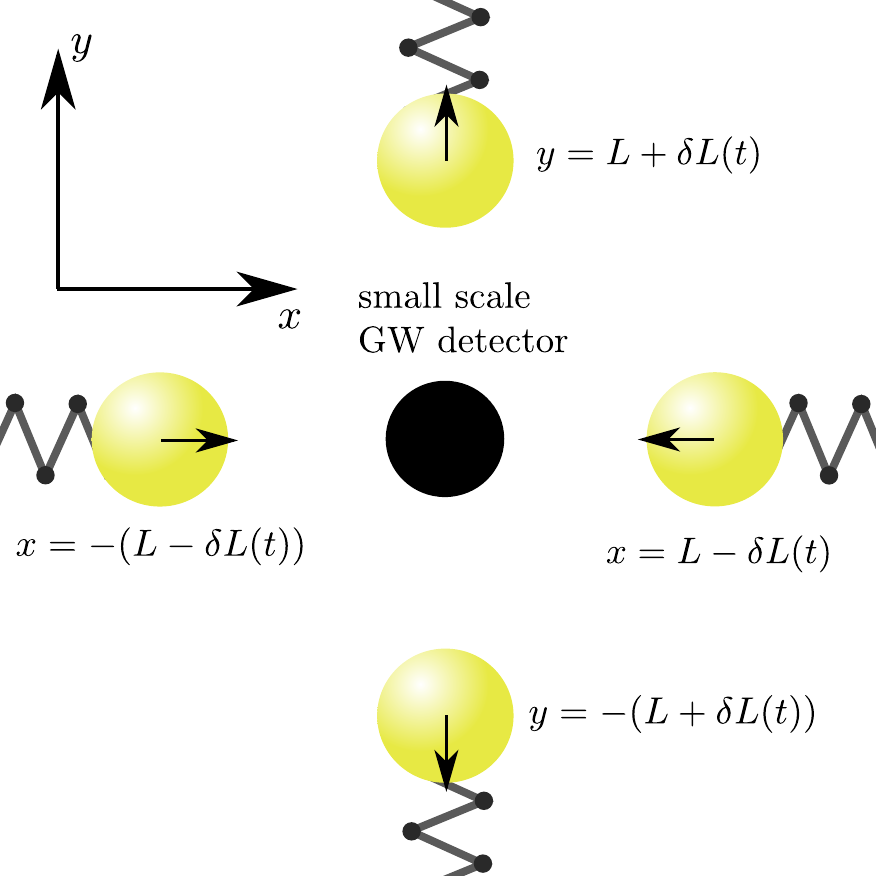}
\caption{\label{fig:spheres} Schematic representation of the setup to mimic the effect of a gravitational wave on a small scale gravitational wave detector. The detector is placed at the center of a distribution of four spheres (possibly hold by levers \cite{Schmoele:2016mde}) that oscillate towards and away from the detector. Opposing spheres have half a period phase shift, while there is a phase shift of quarter of a period between neighboring spheres. A gravitational wave of $\times$-polarization can be mimicked by rotating the setup by $45^\circ$ in the $x$-$y$-plane. Another two masses placed along the $z$-axis above and below the detector are need to cancel constant tidal forces induced by the spheres in the $x$-$y$-plane.}
\end{figure}

The only step that remains in order to mimic a gravitational wave for small scale gravitational wave detectors is the creation of a situation in which $h^\rm{N}_{00}$ matches $h^\rm{w,P}_{00}$. We consider two examples here - a distribution of spheres and a distribution of cylinders. First, let us consider four oscillating spheres of the same mass $M$ placed at the same distance from the detector. Two of these spheres are placed along the $x$-axis on opposite sites of the detector. They oscillate along the $x$-axis in opposite directions. We assume a similar situation for the two other spheres along the $y$-axis (see Fig. \ref{fig:spheres}). Such a situation could be realized by holding the spheres by levers. This would be similar to the experimental proposal described in \cite{Schmoele:2016mde}, where the gravitational field of a single oscillating mass is planned to be measured. The $x$-$y$-plane could be arranged horizontally and the levers could be attached such that the space between the spheres would be empty except for the detector.
In particular, we assume that the spatial positions of the centers of the spheres $(x,y,z)$ are given as $(\pm (L - \delta L(t)),0,0)$ and $(0,\pm (L + \delta L(t)),0)$, where $\delta L(t)=\delta L_0 \cos(\omega t + \phi)$. For the time-time component of the metric perturbation in the Newtonian limit expanded up to quadratic terms in the spatial coordinates, we then obtain 
\begin{equation}\label{eq:hspheres}
	h^\rm{N}_{00}=\frac{2GM}{c^2}\left(\frac{r^2-3z^2}{L^3}+\frac{9}{L^4}(x^2-y^2)\delta L(t)\right)\,,
\end{equation}
where $G$ is Newton's constant and $r=\sqrt{x^2+y^2+z^2}$. The constant term in Eq. (\ref{eq:hspheres}) can be considered as a small offset that has at most a time-independent effect on the detector readout. The constant term vanishes if one places two more spheres at the static spatial positions $(0,0,\pm L)$. Comparison of the oscillating term in Eq. (\ref{eq:hspheres}) with the metric perturbation due to a plane gravitational wave in Eq. (\ref{eq:properdet}) and Eq. (\ref{eq:strains}) leads to the conclusion that we need the conditions $18GM\delta L_0/L^4 = \omega^2 h_{0,+,\omega}/2$ and $\phi=\varphi_+ + \pi$ to be fulfilled to create the effect of the plane gravitational wave of $+$ polarization specified above. Accordingly, we can rotate our whole setup in the $x$-$y$ plane by $45^\circ$ to create the effect of a plane gravitational wave of $\times$ polarization. Since we are working in linearized gravity, we can superimpose the two setups by using 8 spheres (10 to eliminate of the offset) to mimic a gravitational wave of any polarization. Furthermore, we can simulate a general gravitational wave pattern by superimposing oscillations of the masses of different frequencies.

Another possibility to eliminate the constant term in $h^\rm{N}_{00}$ is to replace the spheres with long cylinders oriented in the $z$-direction. Let us assume that the centers of the cylinders follow the same trajectories in the $x$-$y$-plane as we assumed for the centers of the spheres above. Furthermore, let us assume that the cylinders are much longer than the extension of the detector. Then, the Newtonian potential of a single rod can be written as $\Phi^{\rm{rod}}=2\rho A G\ln(R/R_0)$, where $\rho$ is the mass density of the rod, $A$ is its cross section and $R$ is the distance to the center of the rod. $R_0$ is some arbitrary constant of the same dimension as $R$ that is only introduced to obtain a dimensionless expression in the logarithm. It will not contribute to any physical effect here. The time-time component of the metric perturbation for the set of four moving cylinders expanded up to quadratic terms of the spatial coordinates becomes
\begin{equation}\label{eq:hrods}
	h^\rm{N}_{00}=\frac{16\rho AG}{c^2L^3}(x^2-y^2)\delta L(t)\,,
\end{equation}
and we can identify $16\rho AG\delta L_0/L^3 = \omega^2 h_{0,+,\omega}/2$ and $\phi=\varphi_+ + \pi$.

To accurately mimic the component $h^\rm{N}_{00}$ of the metric perturbation due to a gravitational wave, the masses and the detector have to be positioned precisely. Small displacements from the setup described above may lead to a net gravitational force on the center of mass of the detector or additional tidal forces. However, the positioning of the masses and the detector can be achieved with sufficient precision \footnote{Positioning is usually possible with a relative error of, at most, the order of $10^{-3}$ and, in principle, down to an absolute error of $0.1\rm{\mu m}$ given a good reference point using commercially available high accuracy positioners.}, so that the acceleration/additional gravitational effects/tidal forces would be small in comparison to the mimicked effect of a gravitational wave.

Note that the designs of the detectors mentioned in the introduction are so that their detection efficiency is maximal for a certain orientation of the detector with respect to polarization plane of the gravitational wave. In the setup proposed in this article, the polarization plane is fixed ($x$-$y$-plane) and the local detector can always be oriented to maximize detection efficiency.

\section{Signal amplitudes} Let us evaluate the effective signals that one can mimic with spheres and cylinders for some specific experimental parameters. We found that $h_{0,+,\omega}=36MG\delta L_0/L^4\omega^2$ for the spheres and $h_{0,+,\omega} = 32\rho AG\delta L_0/L^3\omega^2$ for the cylinders. Let us assume that the detector systems that we consider have dimensions below the meter scale. Accordingly, we assume that the distance from the detector to the source masses $L$ is of the order of $1\,\rm{m}$. Furthermore, let us assume that the angular frequency $\omega$ of interest is around $2\pi\times 10^3\,\rm{Hz}$.  Let us assume that we are able to move spheres from tungsten or gold of $M=20\,\rm{mg}$ (corresponding to a diameter of about $1\,\rm{mm}$) at this frequency by a distance of about $100\,\rm{\mu m}$. Then, we obtain that we can mimic a strain of the order of $10^{-25}$. For gold or tungsten cylinders with diameters of about $0.5\,\rm{mm}$ as source masses, we find a mimicked strain of the order of $10^{-23}$. For the case of gravitational wave detectors on the centimeter scale like that proposed in \cite{Sabin:2014bua}, the distance between the detector and the source masses $L$ can be reduced to the order of $10\,\rm{cm}$, which leads to mimicked strains of the order of $10^{-21}$ for spheres and $10^{-20}$ for cylinders or strings. 

The strains of gravitational waves expected from persistent cosmic sources are of the order of $10^{-27}$. The mimicked strains are several orders larger and could, therefore, be used as a tool to experimentally test the proposed detector designs. In particular, a strain of the order of $10^{-20}$ is 5 orders of magnitude larger than the strain that is claimed to be detectable with the design proposed in \cite{Sabin:2014bua}. In \cite{Singh:2016xwa}, it is claimed that a sensitivity for strains of the order of $10^{-26}$ could be expected for the first generation detector proposed in the article (denoted as Gen1 in the article). The dimensions of this detector would be about $50\,\rm{cm}$. Therefore, $L$ can be of the order of meters and strains could be mimicked that are  up to 3 orders of magnitude larger than the expected sensitivity of the Gen1 detector in \cite{Singh:2016xwa}.

For frequencies of the order of MHz, larger masses have to be moved to obtain large strains. 
However, besides very extreme sources like galactic center branes, most expected sources of gravitational wave signals of frequencies at or above the MHz regime give rise to strains of the order of $10^{-29}$ and less \cite{Cruise:2012the}. To test appropriate proposals for small scale detectors for the MHz range already during the engineering phase, gravitational waves with strains of the order of $10^{-27}$ could be mimicked with steel strings with diameters of about $100\,\rm{\mu m}$ using oscillation amplitudes of the order of $100\,\rm{\mu m}$ and a distance of $10\,\rm{cm}$ between the sources and the detector.

How vibration isolation of sources and detector can be achieved for experiments with oscillating masses in the milligram range was discussed in \cite{Schmoele:2016mde} and in much detail in the thesis of Jonas Schm\"ole at the University of Vienna \cite{Schmoele:2017diss}. An experimental proposal is presented in which a source mass of about $100\,\rm{mg}$ is moved at a frequency of about $10\,\rm{Hz}$ and its gravitational effect on a test mass is measured. The vibration isolation is implemented through several mechanical vibration isolation stages in the suspension of detector and source. The experiment proposed in \cite{Schmoele:2016mde} is currently set up in the laboratories of Markus Aspelmeyer in Vienna.  Vibration isolation with mechanical isolation stages in the suspension is similar to the techniques developed for vibration isolation of the parts of interferometric gravitational wave detectors like LIGO and Virgo. There, up to 7 individual isolation stages are used in the suspension of the test masses leading to a significant suppression of seismic noise \cite{Ligo:2015adv}. Similar techniques should be used for the test of small scale gravitational wave detectors. A more detailed analysis will be necessary to obtain the precise experimental requirements which will be part of followup project.

\section{Conclusions} We found that the effect of a gravitational wave on a small scale gravitational wave detector of any type mentioned in the introduction can be mimicked by a system of 10 oscillating spheres or a system of 8 oscillating cylinders. In principle, strains up to $10^{-15}$ can be mimicked with masses of the order of $100\,\rm{g}$ oscillating at a frequency of $10^3\,\rm{Hz}$ depending on the size of the detector system. Such strains are larger by a factor $10^6$ than those expected from cosmic sources in this frequency range, in particular, the strongest gravitational wave signal detected by the LIGO-Virgo collaboration \cite{Abbott:2016obs}. Additionally, the orientation of the polarization plane of the mimicked gravitational waves is known and the local detectors can be oriented so that their signal is maximized. Therefore, the setup proposed in this article can be used to test prototypes of small scale detector designs. This would make it possible to test their viability before they even reach the sensitivity needed to detect cosmic gravitational waves, which would require many further years of research and engineering. Viable detector designs could be singled out and developed with concentrated effort. The possibility to mimic gravitational wave signals of any type could also be extremely helpful in the engineering and development stage of the detectors, as problems could be identified and performances could be optimized. The great advantage of the scheme presented here is the possibility of mimicking persistent sources. Most of the small scale detector proposals need large integration times, which makes them more useful for the detection of persistent sources.

The parameters of the setup needed to test a specific small scale gravitational wave detector proposal have to be specified depending on the parameters of the detector design; most importantly the frequency band and the size of the detector. A detailed study for the case of the gravitational wave detector based on a Bose-Einstein condensate proposed in \cite{Sabin:2014bua,Sabin:2015mha} is work in progress.

In our derivation, we assume the Newtonian limit of linearized gravity which requires the assumptions of  non-relativistic motion of the source masses that are used to mimic the effect of a gravitational wave. Furthermore, we assumed that all parts of the detector that are directly interacting with the gravitational field to realize the sensing process are moving with non-relativistic speed. This is not the case for some proposals, such as those presented in \cite{Cruise:2000ane,Cruise:2005aco,Cruise:2006apr}. For these proposals, the additional relativistic effects have to be taken into account.

\begin{acknowledgments} 
We thank Tobias Westphal, Jonas Schm\"ole, Philipp Haslinger and Igor Pikovski for interesting remarks and discussions and Kiri Mochrie for writing assistance. D.R. acknowledges financial support by the Humboldt Foundation. I.F. would like to acknowledge support of the grant "Leaps in cosmology: gravitational wave detection with quantum systems" (No. 58745) from the John Templeton Foundation. 
\end{acknowledgments}

\bibliographystyle{apsrev4-1} 
\bibliography{tests_of_local_detectors}

\end{document}